# Non-invasive improvement of machining by reversible electrochemical doping: a proof of principle with computational modeling

Anastassia Sorkin[1], Yunfa Guo[1], Manabu Ihara[2], Sergei Manzhos[*,2], and Hao Wang[*,1]

[1] *Department of Mechanical Engineering, College of Design and Engineering, National University of Singapore, 9 Engineering Drive 1, Singapore 117575, Singapore*

[2] *School of Materials and Chemical Technology, Tokyo Institute of Technology, Ookayama 2-12-1, Meguro-ku, Tokyo 152-8552 Japan*

Emails: mpewhao@nus.edu.sg (H. Wang); manzhos.s.aa@m.titech.ac.jp (S. Manzhos)

**Abstract**

We propose that the machinability of hard ceramics can be improved by reversible electrochemical doping. On the example of $TiO_2$, we show in a combined density functional theory – molecular dynamics computational study that a small amount of intercalated lithium, which preserves the host structure and can be introduced reversibly, leads to a lowering of the strength of work materials and the cutting force. This is in spite of the fact that there are no significant modifications of the elastic constants at room temperature, i.e. the effect is mostly on plastic properties. This approach is expected to be applicable to a class of ceramics exhibiting similar mechanisms of host-dopant interactions and presents a reversible and non-destructive way of modifying mechanical properties.

## 1 Introduction

Titanium dioxide ($TiO_2$) is a ceramic widely studied for applications in the fields of metal ion batteries [1], solar cells [2], photocatalysis [3], and biomedicine [4], due to its outstanding semiconductive and biocompatible nature. In some applications, mechanical, including plastic and fracture, properties are important, either for machining a ceramic for a given application (e.g. optics [5]) or because of conditions of use (e.g. implants operated under mechanical stress [6]). Ultra-precision mechanical machining is required to obtain complex geometries with superior surface finishing for engineering applications of ceramics such as electronics and optics industries. However, many ceramics including $TiO_2$ considered here are brittle materials and tend to deform by catastrophic fracture with the generation of microcracks on the surface and subsurface. The intrinsic brittleness of $TiO_2$ makes it rather challenging to manufacture using ultra-precision machining technology. Ductile-mode cutting, a method of material removal by plastic deformation, has been successfully applied in the ultra-precision machining of brittle materials to generate crack-free surfaces. Traditionally, ductile-mode cutting is introduced by decreasing the undeformed chip thickness to the submicrometric range for ductile–brittle transition, which significantly restricts the manufacturing efficiency and generates high machining costs [7].

Due to the poor productivity of brittle materials under conventional ductile-mode cutting, efforts have been made in the academic and industrial communities to extend the ductile–brittle transition range, which includes thermal-assisted machining [8], ultrasonic vibration-assisted machining [9], and magnetic field-assisted machining [10]. Moreover, surface modification of work materials has also revealed its potential to increase the critical undeformed chip thickness for ductile-brittle transition in machining brittle materials. A work using the application of the solidified coating on the workpiece surface by Lee et al. [11] showed an increase in critical undeformed chip thickness during micro-cutting of $CaF_2$ single crystal. Additional reaction stress was believed to be induced on the shear plane of the work materials due to the non-deformability of the solidified coating, which weakened the effective tensile stress on cleavage planes and thus suppressed the crack growth to facilitate the ductile-brittle transition [12]. However, cutting with solidified coating increases the cutting energy due to the unavoidable removal of the applied coating. Ion implantation on the surface of work materials is another method for surface modification [13,14]. To et al. [14] implanted hydrogen ions into a silicon surface to enhance the critical chip thickness during micro-cutting of silicon, which would induce irreversible secondary defect generation such as sputtering of the work material, lattice distortion, and amorphous layer formation. Therefore, it is essential to find a new surface modification technology to break through the technical barriers of solidified coating and hydrogen ion irradiation.

Electrochemical doping of oxides is expected and in certain cases has been demonstrated to lead to mechanical softening. We are interested in the non-destructive and reversible modification of mechanical properties by interstitial doping rather than substitutional doping to facilitate machining without affecting the mechanical properties of host's structure after machining. It can be achieved through small-concentration interstitial doping which does not affect crystal structure and crystal site identity beyond a certain degree of relaxation of neighboring atoms via electrochemical doping. This is in particular the case for the considered here lithium intercalation into titania. It has been widely studied for the applications in high-rate lithium-ion batteries (LIB) [15–18]. The present study will employ the lithium intercalation into titania as a case to investigate the potential of low-concentration electrochemical doping for the improvement of mechanical properties and machining performance of ceramics. It is known that the crystal structure of major titania phases (anatase, rutile, and bronze) can accommodate at least up to 0.25 Li per formula unit without changes in the crystal structure [19]. Therefore, this work will be deploying Li concentrations not exceeding 0.25 Li per formula unit to preserve the host's structure and achieve the aim of reversible electrochemical doping. Li interstitials are also highly mobile in $TiO_2$ with some reported diffusion barriers computed by density functional theory (DFT) as low as 0.0004 eV [19]. The latter value, implying practically barrierless diffusion, was computed for Li diffusion along the c-axis in rutile, the phase considered here [19]. The often-used anatase phase also exhibits relatively benign barriers on the order of 0.5 eV [19–21]. This forms the basis for the use of Li-$TiO_2$ system as high-rate anodes for LIB, which have recently been commercialized by Toshiba [22]. Another reason besides the low diffusion barrier is the intercalation potential, of about 1.75 V vs Li/$Li^+$ at low Li concentrations, which allows avoiding the formation of solid-electrolyte interphase via reductive decomposition of common organic (carbonate based) electrolytes. This implies that both the bulk and the surface of titania can remain stable under (de-)intercalation.

Mechanistically, it is expected that the intercalation of species easily forming ionic compounds, such as Li, Na, Mg, and Al, has the potential to soften a semiconductor ceramic. The valence electrons of the ionized intercalating species occupy states near the bottom of the conduction band or bandgap states (such as $Ti^{3+}$ states formed by Ti atoms reduction by the valence electrons of Li inserted into $TiO_2$ [19,23,24]). Those states are non-bonding [25,26]. Such mechanistic insight is easily obtainable with DFT. The necessarily small-scale DFT simulations suggest but do not necessarily predict that semiconducting ceramics could be made less brittle by intercalation of small amounts of easily ionized atoms. While elastic properties are easily computable at the DFT level, plastic response requires large-scale models. To confirm reliably the effect on machining, larger scale simulations of plastic properties are required which on one hand need to treat plastic effects directly and on the other hand include the mechanistic information consistent with the results of ab initio simulations. This can be done with force field molecular dynamics employing a force field giving justice to key mechanistic details. The electronic mechanism described above may well lead to softening but it acts in combination with the effect of strain induced by the insertion of the dopant which can lead to hardening [27–29]. Depending on the type of inserted atom, the strain component can even dominate the defect formation energy, but this component is the smallest with Li due to its small ionic radius and charge. Lithium is therefore most likely to lead to substantial softening. A direct simulation of the change in mechanical properties in the plastic regime would ascertain the combined effect on the mechanical properties of these factors.

This work offers a multi-scale numerical investigation including small scale density functional theory calculations and large scale molecular dynamics simulations to comprehensively reveal the effect of small-concentration lithiation on mechanical properties in the plastic and fracture regimes of rutile $TiO_2$. We modelled this system in the elastic regime with both DFT and a force field accounting for the key mechanism of lithium-titania interaction, i.e. the oxidation of Li and formation of $Ti^{3+}$ (formal oxidation state) by reduction of (formal) $Ti^{4+}$ with Li valence electrons. In this way, we benchmark the force field and ascertain the mechanism of lithiation. Large-scale molecular dynamics simulations of plastic properties, including stress-strain curves up to fracture and modelling of cutting experiments, were conducted. The lithiation-induced changes in stress-strain curves, chip morphology, and cutting force were analyzed. We show that the material is softened and cutting force is decreased in the presence of intercalated Li. As similar mechanisms are active in different ceramics, we expect this approach to be useful to facilitate machining with non-destructive reversible modifications of composition.

## 2  Methods

### 2.1  Ab initio calculations

DFT (density functional theory [30]) calculations were performed in SIESTA code [31] using PBE exchange-correlation functional [32]. Troullier-Martins pseudopotentials were used for core electrons [33,34]. In the case of Ti, 4s and 3d electrons were included in the valence configuration, and 2s and 2p electrons were included for O and 2s for Li. A smaller-core pseudopotential for Ti was tried but did not significantly change the results. A DZP (double-z polarized) basis set was used and generated with the option PAO.EnergyShift = 0.001 Ry to improve the completeness of the basis. Electron density was expanded with the plane wave cutoff of 200 Ry. Electronic occupations were smeared with a Fermi-Dirac function with an electronic temperature of 500 K. The rutile TiO$_2$ cell used in the ab initio calculations consisted of 2×2×4 primitive cells (about 9×9×12 Å), which is 96 atoms (32 of them are Ti atoms and 64 are O atoms). The Li atom was inserted into the octahedral site which previously was shown to be the most energetically favorable [21]. This corresponds to a Li$_{0.03125}$TiO$_2$ stoichiometry. The Brillouin zone was sampled with 3×3×3 Monkhorst-Pack points. Structures were relaxed until atomic forces were below 0.005 eV/Å and pressure below 0.005 GPa. In order to correctly reproduce the electronic structure of Li-doped TiO$_2$, in particular the localized gap states corresponding to the formation of Ti$^{3+}$ (formal oxidation state) [20,35], we applied a Dudarev +U correction with $U = 3$ eV applied to Ti $d$ states (PBE + U) [36].

## 2.2 Molecular dynamics simulations

The molecular dynamics simulations were carried out using Large-scale Atomic / Molecular Massively Parallel Simulator (LAMMPS) code [37]. In this work, the short-range interactions were described using a Buckingham potential with a long-range Coulomb term with coefficients proposed by Matsui and Akaogi (see Table 1) [38]. The Buckingham potential is expressed as follows:

$$U(r_{rj}) = A_{ij} exp\left(-\frac{r_{ij}}{\rho_{ij}}\right) - \frac{C_{ij}}{r_{ij}^6} + \frac{q_i q_j}{r_{ij}} \qquad (1)$$

**Table 1.** Parameters for the Buckingham potential used in this work.

| ion pair (ij) | $A_{ij}$ (eV) | $\rho_{ij}$ (Å) | $C_{ij}$ (eV*Å$^6$) |
|---|---|---|---|
| Ti-Ti | 31120.528 | 0.154 | 5.25 |
| Ti-O | 16957.710 | 0.194 | 12.59 |
| O-O | 11782.884 | 0.234 | 30.22 |
| Ti-Li | 33089.570 | 0.127 | 0.00 |
| O-Li | 15465.549 | 0.167 | 0.00 |
| Li-Li | 38533.955 | 0.100 | 0.00 |

Following Kerisit et al. [39,40], we have kept the partial charges ($q$) of oxygen and titanium atoms equal to these suggested by Matsui and Akaogi, and the partial charge of Li was determined by using the Li$_2$O stoichiometry. These values are −1.098, +2.196, and +0.549 |$e$| for formal O$^{2-}$, Ti$^{4+}$, and Li$^+$, respectively. According to the DFT calculations, small electron polarons are formed at titanium lattice sites, formally reducing them to Ti$^{3+}$ cations (in contrast to other Ti atoms which are formally Ti$^{4+}$). There is one Ti$^{3+}$ atom for each Li atom. Ti$^{3+}$ atoms were created by assigning a charge +1.647 to select Ti atoms to compensate for the charge of Li. Note that while these charges are different from the formal oxidation states of the respective anions, so are actual atomic charges due to the partially covalent character of the bonds and charge self-regulation in transition metals [15,41].

The samples were relaxed at zero pressure at 300 K during 10 ps. The lengths of the unit cell in the relaxed samples can be found in Table 2. As in the DFT calculation, the $L_x$ is not equal to $L_y$ for $Li_{0.03125}TiO_2$. To compute the plastic response, molecular dynamics simulations were performed with an NPT ensemble with the Nose-Hoover thermostat at 300 K and a barostat, both with a reaction time of 0.05 ps. The simulation cell was stretched in *x, y*, or *z*-direction, (parallel to a, b and c axis of the rutile orthorombic cell, respectively) with a constant strain rate of 0.001 1/ps, in such a way that the box length *L* as a function of time will change as

$$L(t) = L_0(1 + erate \times dt) \qquad (2)$$

where $L_0$ is initial length of the sample (optimized geometry), *erate* is a strain rate and *dt* is a timestep equal to 0.2 fs. The pressure was kept equal to zero in two perpendicular directions by the barostat. The structures were visualized with the VESTA visualization program [42].

## 3  Results and discussion

### 3.1  Elastic properties

The DFT+U-optimized $Li_{0.03125}TiO_2$ cell is shown in Figure 1. It can be seen that Li intercalation leads to a slight deformation of the original $TiO_2$ cell. The vector angles remain very close to 90°; however, in contrast to $TiO_2$, the length of the vector *x* is not equal to the length of the vector *y*. The elastic modulus was evaluated in the DFT calculations by expansion of the linear cell size within 1%. The computed Exx = Eyy and Ezz components of the elastic modulus tensor for $TiO_2$ are 250.0 and 416.5 GPa, respectively, which is somewhat smaller than the experimental values reported in Ref. [43] (268 and 484 GPa). We note that experimental values are not at the single-crystal level and therefore an exact match is not expected. Table 2 lists the calculated lattice constants (in Å) and elastic moduli (in GPa) for pristine $TiO_2$ rutile and lithiated rutile $Li_{0.03125}TiO_2$ with PBE+U as well as the force field. It can be seen that intercalation of Li at such a small concentration does not show any noticeable effect on the elastic constants. This may have to do with the fact that Li valence electrons occupy nonbonding rather than antibonding orbitals of the host. In agreement with previous calculations [24], Li doping was found to be energetically favorable with formation energy of -1.62 eV vs Li bulk and TiO$_2$ bulk (compare to -1.66 eV from Ref. [24]).

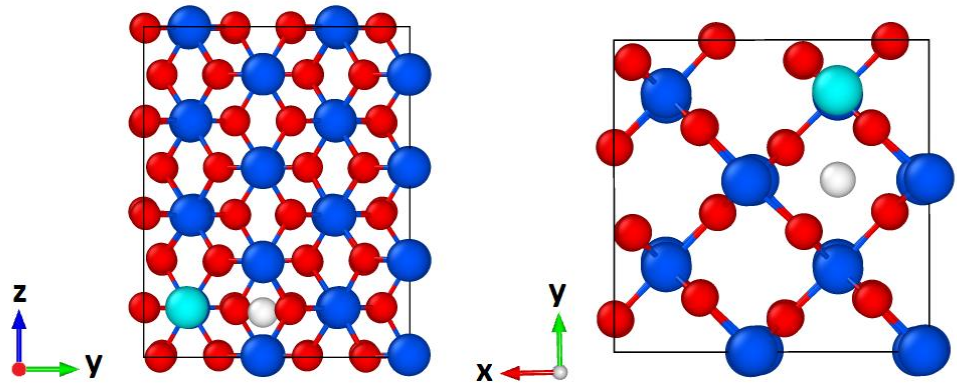

**Figure 1**. The side view (a) and the top view (b) of TiO$_2$ rutile 96-atoms cell with a Li atom optimized with SIESTA. Atom color scheme: Ti$^{4+}$ dark blue, Ti$^{3+}$ light blue, O – red, Li - white.

**Table 2.** Lattice constants (in Å) and elastic moduli (in GPa) for pristine TiO$_2$ rutile and for lithiated rutile Li$_{0.03125}$TiO$_2$ calculated in this work with PBE+U as well as the force field. (For Li$_{0.03125}$TiO$_2$ effective lattice constants of the putative titania unit cell are calculated from the 96-atom simulation cell). a-Ref. [44], b-Ref. [45], c-Ref. [43]. and d-Ref. [46].

|  | TiO$_2$ | | | Li$_{0.03125}$TiO$_2$ | |
|---|---|---|---|---|---|
|  | PBE+U | Force field | exp. | PBE+U | Force field |
| $L_x$ | 4.621 | 4.493 | 4.58$^a$ 4.584$^b$ | 4.634 | 4.484 |
| $L_y$ | 4.621 | 4.493 | 4.58$^a$ 4.584$^b$ | 4.641 | 4.515 |
| $L_z$ | 3.023 | 3.009 | 2.95$^a$ 2.953$^b$ | 3.025 | 3.005 |
| $E_{xx}$ | 250.0 | 253.6 | 268.0$^c$ 292.0$^d$ | 260.0 | 256.5 |
| $E_{yy}$ | 250.0 | 253.7 | 268.0$^c$ 292.0$^d$ | 259.9 | 264.8 |
| $E_{zz}$ | 416.5 | 538.6 | 484.0$^c$ 471.0$^d$ | 410.0 | 533.4 |

The $E_{xx}$ and $E_{yy}$ components of the elastic modulus for TiO$_2$ rutile calculated with the Buckingham force field are very close to those calculated with DFT, while the $E_{zz}$ component is bigger than DFT results as well as the experimental value, however it is in the same range of deviation from the experimental value as the DFT result. In presence of lithium the results of calculations with the force field slightly change the symmetry between $x$- and $y$- edges of the lattice, but do not change the vector angles, in agreement with the DFT results. In agreement with the DFT results, the intercalation of the small amount of lithium do not cause a noticeable change in elastic constant: both DFT and MD show only a slight increase of the $E_{xx}$ and $E_{yy}$ and slight decrease of the $E_{zz}$, indicating no significant effect in the linear regime. These results also ascertain the qualitatively correct behaviour of the force field, which is used in large scale MD simulations in the following.

Figure 2 shows the calculated electronic density of states for the $Li_{0.03125}TiO_2$. A defect state in the band gap lies approximately 0.7 eV from the bottom of the conduction band. The plot of Ti $3d$ densities of states confirms that these defect states are occupied by Ti $3d$ states. The Mulliken population analysis also clearly shows the reduction of one Ti atom. The Li's $s$ shell is ionized, and its Mulliken population is 0.19 only, while the reduced Ti atom has its $d$ shell Mulliken population increase by about 0.5 vs other Ti atoms and vs pure $TiO_2$. That is, the calculation correctly reproduces the well-known electronic structure features of this system, in particular the appearance of $Ti^{3+}$ which we model explicitly in the force field simulations of plasticity. (Note that Mulliken populations are well-known to imperfectly render charge changes so an exact change of the relevant populations by $1e$ is not expected, especially with rather broad basis functions used here.)

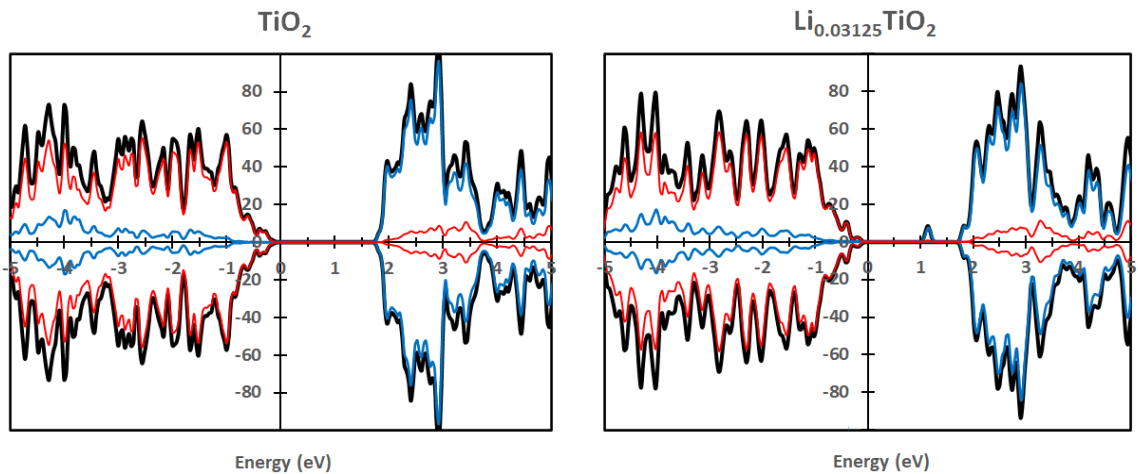

**Figure 2**. Total (black) and partial (O($p$) – red, Ti($d$) - blue) densities of states for pristine $TiO_2$ rutile and $TiO_2$ rutile doped with 1 atom of Li. The DOS of $Li_{0.03125}TiO_2$ shows the standalone state inside the band gap at near 1 eV corresponding to the formation of $Ti^{3+}$. The origin of the energy axis is set at the top of the valence band.

### 3.2 Plasticity

In the model setup, one to five Li atoms were respectively inserted into a 48-atom $TiO_2$ rutile cell (containing 2×2×2 primitive cells) by keeping the distance between the Li atoms as large as possible (considering Coulombic repulsion between Li$^+$) to form the unit cells of Li$_{0.0625}$TiO$_2$, Li$_{0.125}$TiO$_2$, Li$_{0.1875}$TiO$_2$ and Li$_{0.25}$TiO$_2$ compounds. An example of the simulation cell containing 4 Li atoms is shown in Figure 3. These cells were then replicated 5×5×5 times (6500 atoms including Li), relaxed during 10 ps, and stretched to obtain the stress-strain curves with molecular dynamics simulations.

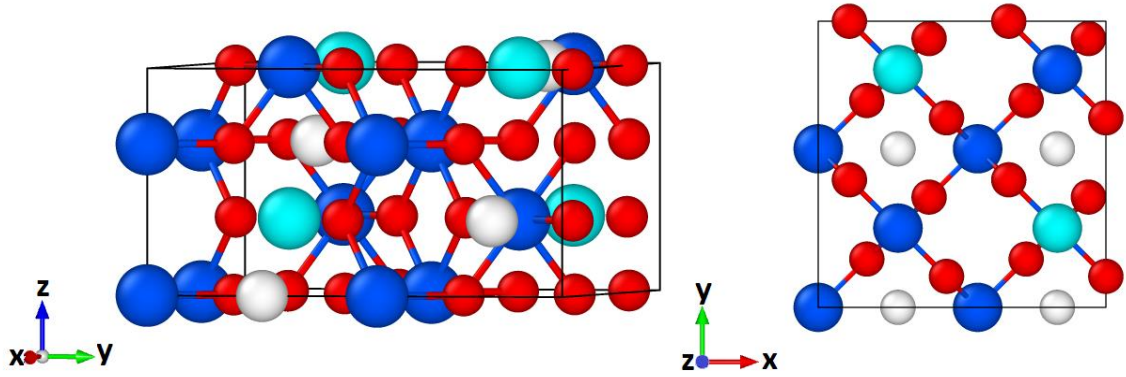

**Figure 3**. The side view (left) and the top view (right) of the cell of Li$_{0.25}$TiO$_2$ used in this work. The dark blue atoms are Ti$^{4+}$, the light blue atoms are Ti$^{3+}$, the red atoms are O, while the small white atoms are Li.

**Table 3.** Lattice parameters (in Å) of lithiated TiO$_2$ with different Li concentrations, as well as diagonal components of the elastic tensor $G$ (in GPa) and the fracture stresses $F$ (in GPa) found from the stress-strain curves. Lattice parameters were computed for putative titania unit cell from the lithiated cells described in the text.

|  | TiO$_2$ | Li$_{0.0625}$TiO$_2$ | Li$_{0.125}$TiO$_2$ | Li$_{0.1875}$TiO$_2$ | Li$_{0.25}$TiO$_2$ |
|---|---|---|---|---|---|
| $L_x$ | 4.503 | 4.506 | 4.532 | 4.545 | 4.549 |
| $L_y$ | 4.503 | 4.528 | 4.525 | 4.551 | 4.605 |
| $L_z$ | 3.018 | 3.013 | 3.011 | 3.008 | 2.995 |
| $E_{xx}$ | 251.2 | 251.7 | 250.4 | 248.8 | 251.1 |
| $E_{yy}$ | 251.7 | 250.3 | 249.0 | 262.8 | 261.5 |
| $E_{zz}$ | 536.9 | 526.2 | 523.3 | 499.2 | 463.8 |
| $F_{xx}$ | 52.7 | 47.1 | 45.0 | 44.7 | 40.1 |
| $F_{xx}$ | 52.6 | 49.0 | 44.1 | 45.8 | 42.3 |
| $F_{xx}$ | 60.1 | 50.3 | 46.8 | 44.2 | 37.5 |

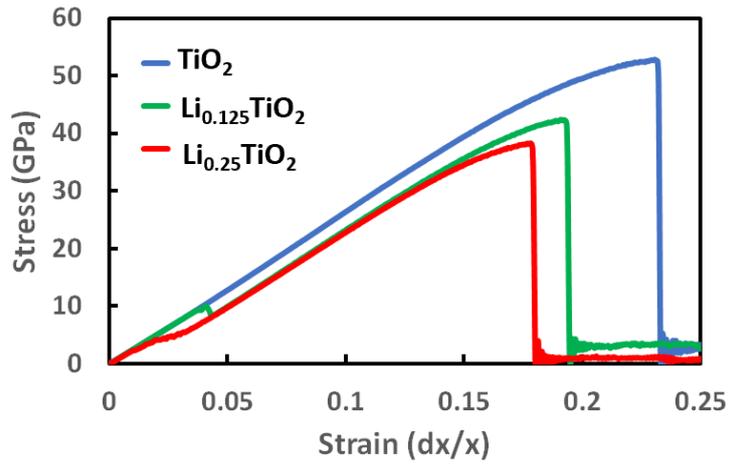

x-direction

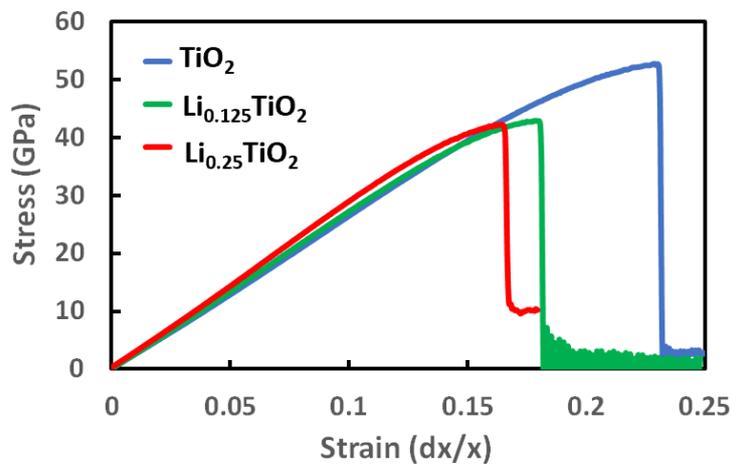

y-direction

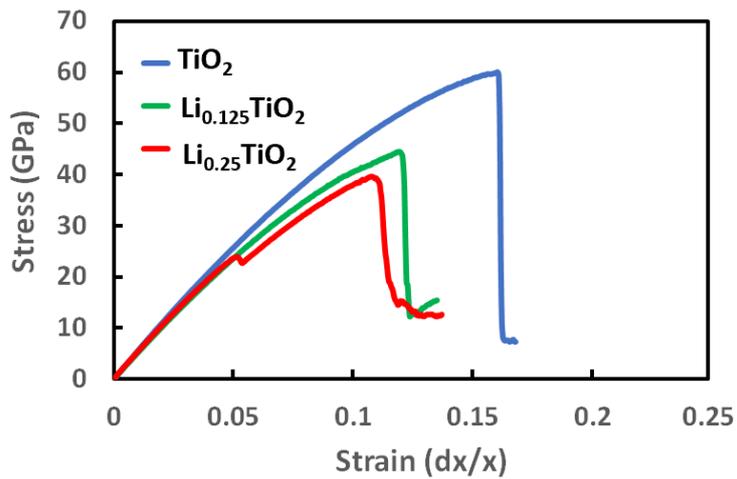

z-direction

**Figure 4**. The strain-stress curves for lithiated $TiO_2$ with different Li concentration.

The geometries of the relaxed samples are described in Table 3. For most of the samples, the lattice constant is increasing with Li concentration in the *x* and *y* directions and is slightly decreasing in the *z* direction. However, the geometry parameters depend quite strongly on the configuration of Li in the sample and can vary in the samples with the same Li concentration but different configurations. In our simulations, we put Li in the farthest positions from each other as a guess of the lowest energy configuration, based on Coulombic repulsions between Li cations. This is sufficient for the purpose of this study.

Figure 4 shows the stress-strain curves of $TiO_2$, $Li_{0.125}TiO_2$, and $Li_{0.25}TiO_2$, in all three directions under the constant strain rate of 0.001 1/ps. All the curves clearly show the elastic regime up to 2-3 % of strain. The Young moduli $E_{xx}$, $E_{yy}$, and $E_{zz}$ can be derived from the slope of the stress-strain curves, and lattice parameters together with the fracture stress $F_{xx}$, $F_{yy}$, and $F_{zz}$ were computed for putative titania unit cell from the lithiated cells described in the text. Table 3 lists the calculated lattice parameters (in Å), diagonal components of the elastic tensor $G$ (in GPa), and fracture stresses $F$ (in GPa) of lithiated $TiO_2$ with different Li concentrations. Our result for the $TiO_2$ fracture strain in the *x* direction (23.5%) is in good agreement with the molecular dynamics calculation from Ref. [47], which gave a fracture at 20%.

The Young modulus does not change significantly in the *x* direction and even slightly increases in the *y* direction with increasing Li concentration in agreement with the DFT results. However, it steadily decreases in the *z* direction, which is probably related to the higher mobility of Li atoms in the *z* direction than in the *x* and *y* directions [19–21]. Importantly, the fracture strain and stress show a clear trend to decrease with Li concentration in all three directions, demonstrating a significant effect of lithiation on the plastic response even as there is no significant effect on the elastic properties. It may be attributed to more easily formed defects with lithiation which can promote the fracture. It is also important to note that the curves in Figure 4 for $Li_{0.125}TiO_2$ in the *x* direction and for $Li_{0.25}TiO_2$ in *z* direction have a small step at about 4-5% strain. The visualization shows a change of many Li atoms' positions when the step is formed along the *z* direction while the $TiO_2$ lattice does not experience any phase transition. Our initial choice of Li positions at maximum distance was based on minimizing Coulombic repulsion at 0% of strain. But such a configuration might not possess the lowest energy in the cell with the distorted stretched geometry at 4-5% of strain. As Li diffusion barrier in $TiO_2$ is known to be low, such relaxations are therefore not surprising. The collective movement of Li atoms at about 4-5 % of strain slightly releases the stress. Our structures are however sufficient for the purpose of the study.

While the model with 6500 atoms allows seeing qualitatively the influence of Li intercalation on plastic properties, it is still quite small from the point of view of microstructure. In our two last sets of calculations, we created a 48-atom sample of $TiO_2$ with 4 Li atoms ($Li_{0.25}TiO_2$) and replicated it 10×10×10 times and 25×25×25 times. The results for 10×10×10 (52000 atoms) and 25×25×25 supercells (812500 atoms) of $Li_{0.25}TiO_2$ are described in Table 4 in comparison with the previous sample (5×5×5, or 6500 atoms). The stress-strain curves in the *x* direction for the samples of pure $TiO_2$ and $Li_{0.25}TiO_2$ with 5×5×5, 10×10×10, and 25×25×25 supercells are plotted in Figure 5 to reveal the influence of the sample size on its mechanical properties. For $TiO_2$, the graphs do not show a significant deviation between the big and the small samples. Table 4 shows that the elastic constant for $Li_{0.25}TiO_2$ is decreasing as the size of the sample increasing. However, the step on the stress-strain curve which occurs at about 4-5 % of strain is clearly deeper on the smaller samples. After the step the graphs for different cell sizes continue very close to each other up to 15 % of strain. For $Li_{0.25}TiO_2$, the fracture stress for the 25×25×25 samples is clearly smaller than for the 5×5×5 and 10×10×10 models. These results also demonstrate that a sufficiently large-scale model is required to realistically model the generation and propagation of defects and dislocations in the plastic regime.

**Table 4.** The parameters of the unit cell (in Å) for $TiO_2$ and $Li_{0.25}TiO_2$, used in this work as well as diagonal components of the elastic tensor (in GPa) and the fracture stress (in GPa) found from the stress-strain curves computed with simulation cells of different sizes. Lattice parameters were computed for putative titania unit cell from the lithiated cells described in the text.

|   | $TiO_2$ | | | $Li_{0.25}TiO_2$ | | |
|---|---|---|---|---|---|---|
|   | 6000 atoms | 48000 atoms | 750000 atoms | 6500 atoms | 52000 atoms | 812500 atoms |
| $L_x$ | 4.503 | 4.502 | 4.502 | 4.549 | 4.571 | 4.58 |
| $L_y$ | 4.503 | 4.501 | 4.501 | 4.605 | 4.634 | 4.611 |
| $L_z$ | 3.018 | 3.017 | 3.017 | 2.995 | 3.007 | 3.015 |
| $E_{xx}$ | 251.2 | 248.9 | 249.1 | 251.1 | 239.8 | 217.3 |
| $E_{yy}$ | 251.7 | 253.6 | 248.6 | 261.5 | 253.7 | 219.2 |
| $E_{zz}$ | 536.9 | 535.9 | 536.1 | 463.8 | 465.2 | 422.4 |
| $F_{xx}$ | 52.7 | 52.5 | 52.4 | 40.1 | 39.1 | 35.6 |
| $F_{yy}$ | 52.6 | 52.5 | 52.2 | 42.3 | 37.9 | 36.2 |
| $F_{zz}$ | 60.1 | 59.5 | 59.4 | 37.5 | 40.1 | 30.6 |

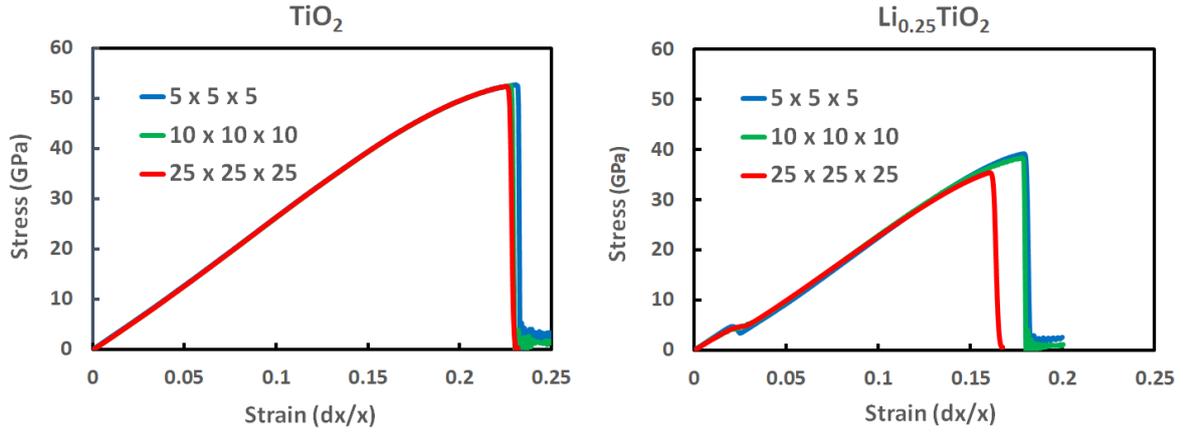

**Figure 5**. Influence of the size of the sample on its mechanical properties.

Most experimentally derived values of elastic constants were obtained in a wide range of temperatures up to 1800 K [43,46]. Those results show that the elastic constant is decreasing with increasing temperature. We carried out the calculation of the stress-strain curve of rutile and lithiated rutile at two additional temperatures of 1000 K and 1500 K with a sample of about 1 million particles. Table 5 and Figure 6 illustrate the influence of the temperature of $TiO_2$, $Li_{0.125}TiO_2$ and $Li_{0.25}TiO_2$ on their mechanical properties. Note that even at 1500 K we did not observe the phase transition to the cubic phase [48], because the pressure during the simulation is negative. In agreement with the experimental measurements, the elastic modulus of rutile as well as the fracture stress decrease with increasing temperature. The same effect takes place for $Li_{0.125}TiO_2$ and $Li_{0.25}TiO_2$. The elastic modulus and the fracture stress for lithiated rutile are smaller than that for pristine rutile at higher temperatures. The results are shown for the *x* and *z* directions only, because the values of the elastic modulus and fracture stress in the *y* direction are very similar to those in the *x* direction. The fracture stress $F_{xx}$ for rutile is decreased by about 28% when the temperature rises from 300 K to 1500 K, while for $Li_{0.25}TiO_2$ the decrease is 46%. Note that for the 25×25×25 sample at 300 K the elastic modulus decreases with increasing Li concentration in the *x* (and *y*) directions, in contrast to the 5×5×5 sample, where the elastic modulus did not show a significant change. At 1000 and 1500 K it also shows a decrease when the concentration of Li increases. The fracture stress decreases with Li concentration at all temperatures.

**Table 5**. The parameters of the unit cell (in Å) for $TiO_2$, $Li_{0.125}TiO_2$ and $Li_{0.25}TiO_2$, used in this work as well as diagonal components of the elastic tensor (in GPa) and the fracture stress (in GPa) found from the stress-strain curves computed with the sample of 25×25×25 times replicated 48-atoms unit cell at different temperatures.

|  | $TiO_2$ | | | $Li_{0.125}TiO_2$ | | | $Li_{0.25}TiO_2$ | | |
| --- | --- | --- | --- | --- | --- | --- | --- | --- | --- |
|  | 300 K | 1000 K | 1500 K | 300 K | 1000 K | 1500 K | 300 K | 1000 K | 1500 K |
| $L_x$ | 4.502 | 4.522 | 4.538 | 4.548 | 4.576 | 4.599 | 4.590 | 4.630 | 4.662 |
| $L_y$ | 4.501 | 4.523 | 4.539 | 4.548 | 4.576 | 4.499 | 4.611 | 4.629 | 4.663 |
| $L_z$ | 3.017 | 3.041 | 3.061 | 3.019 | 3.046 | 3.068 | 3.020 | 3.045 | 3.068 |
| $E_{xx}$ | 249.1 | 237.6 | 227.0 | 250.9 | 232.4 | 214.3 | 217.3 | 204.6 | 202.5 |
| $E_{yy}$ | 248.6 | 237.7 | 227.2 | 245.9 | 230.6 | 218.7 | 219.5 | 211.0 | 203.8 |
| $E_{zz}$ | 536.1 | 484.1 | 445.2 | 489.6 | 436.3 | 390.3 | 422.4 | 371.2 | 327.9 |
| $F_{xx}$ | 52.4 | 43.3 | 37.8 | 41.8 | 32.6 | 26.8 | 35.6 | 26.4 | 19.2 |
| $F_{zz}$ | 52.2 | 43.4 | 38.0 | 42.8 | 33.3 | 26.8 | 36.2 | 26.0 | 19.0 |
| $F_{zz}$ | 59.4 | 45.2 | 37.8 | 43.3 | 32.8 | 26.6 | 30.6 | 24.3 | 19.7 |

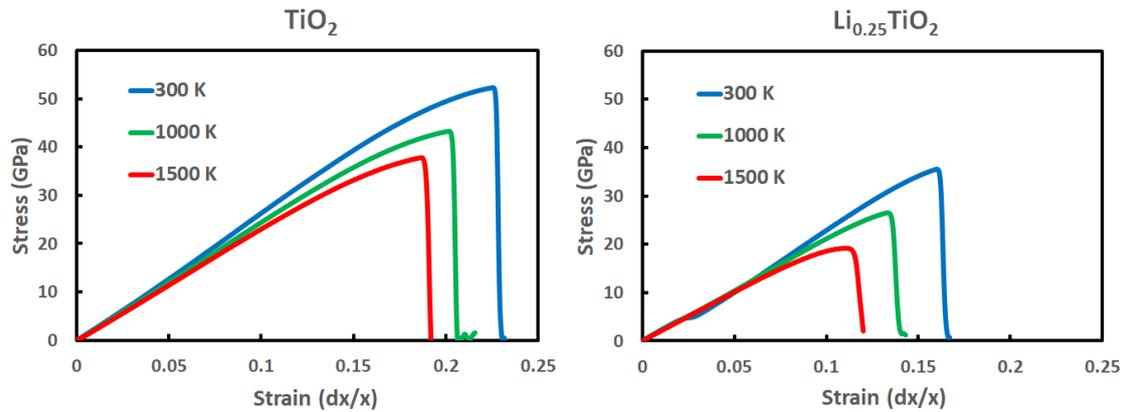

**Figure 6**. Influence of the temperature of the samples of $TiO_2$, $Li_{0.125}TiO_2$ and $Li_{0.25}TiO_2$ on its mechanical properties.

### 3.3 Cutting

A sample of rutile $TiO_2$ with the size of 360×135×175 Å (corresponding to 40×15×30 of 48-atoms unit cell, or 864000 atoms) was employed in the cutting simulation. Two and four Li atoms were respectively inserted into the initial 48-atoms unit cell of $TiO_2$ to simulate the cutting of $Li_{0.125}TiO_2$ and $Li_{0.25}TiO_2$, and the final samples of $Li_{0.125}TiO_2$ and $Li_{0.25}TiO_2$ contained 36000 and 72000 Li atoms. The samples were relaxed at 300 K during 10 ps with periodic boundary conditions. Then the periodic boundary conditions in the *x*- and *z*-direction were lifted, the layers of thickness of 10 Å on the bottom and the left side were frozen and the samples were relaxed again. The cutting tool was a rod in the *y* direction made from a diamond with a triangle cross-section where the lower edge was at the angle of 30º to the *x* axis, the upper edge was at the angle of 60º to the *x* axis, and the right edge was parallel to the *yz* plane (see Figure 7). The length of the cutting tool in the *x* direction was 75 Å. The length of the road in the *y* direction was 135 Å (equal to the length of the sample in the *y* direction). Carbon atoms of the cutting tool interacted with each other by the Tersoff potential [49]. An interaction of carbon atoms with Ti, O, and Li was described with Lennard-Jones potentials with parameters from the Lorenz-Berthelot combining rule. The parameters are listed in Table 6.

**Table 6**. The parameters of the Lennard-Jones potential used in this work for interactions with the cutting tool in cutting simulations.

|        | $\epsilon$ (eV) | $\sigma$ (Å) | $r_{cut}$ (Å) |
|--------|-----------------|--------------|---------------|
| Ti - C | 0.165           | 2.44         | 10            |
| O - C  | 0.0222          | 3.7127       | 10            |
| Li - C | 0.004985        | 2.6325       | 10            |

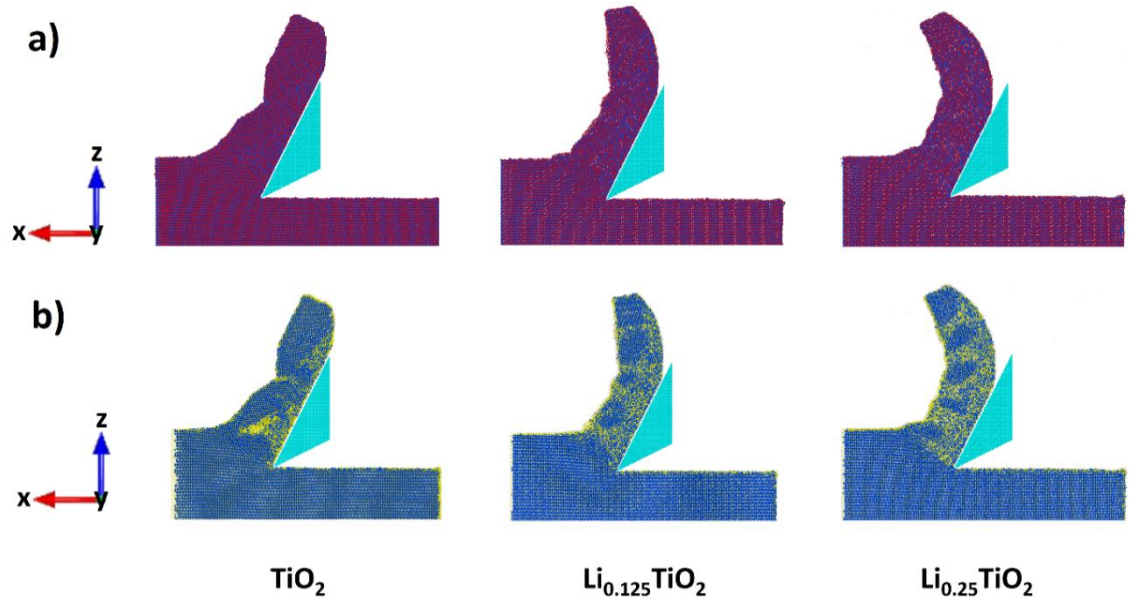

**Figure 7**. Cutting of the rutile and lithiated rutile. Light blue triangle is a diamond tool. a) a snapshot the simulation. Titanium atoms are in blue, oxygen in red (so that $TiO_2$ looks purple) while Li atoms are white. b) Titanium atoms only: blue balls are titanium atoms.

The diamond tool was moved along the *x* axis with a velocity of 2.5 Å/ps (see Figure 7). The simulation was carried out with NVE ensemble at an initial temperature of 300 K. Periodic boundary conditions were applied in the *y* direction (perpendicular to the page). The lowest and the rightmost titania layers with 10 Å of the edges were kept frozen. The adjacent bottom and right layers of 10 Å were kept at 300 K by rescaling the velocities of each atom. This was done to provide the heating outflow. Thus the simulation box was divided into three parts: the frozen atoms, the atoms kept at constant temperature and volume (NVT), and the biggest part of the sample was treated as an NVE ensemble. This approach has been used in previous simulations of cutting [50,51]. The force in the *x* direction acting on all atoms of the cutting tool was measured as the cutting force. A snapshot of the cutting process is represented in Figure 7a.

Figure 8 shows the cutting force measured every 2 ps (or each 5 Å of the tool path) in the cutting of $TiO_2$, $Li_{0.125}TiO_2$, and $Li_{0.25}TiO_2$. It can be seen from the dashed lines in Figure 8 that the average value of cutting force decreases from 0.503 meV/Å to 0.449 meV/Å to 0.411 meV/Å with increasing Li concentration. The reduction of the cutting force is consistent with the reduction in toughness observed in section 3.2. The fracture stress of lithiated $TiO_2$ is smaller than pristine $TiO_2$, indicating that the crystal of lithiated $TiO_2$ is easier to break and then release the stress in shear band suppressing the fluctuation of cutting force. Moreover, the cutting force also exhibits fluctuations with the cutting distance for all samples, but the amplitude of the fluctuations is significantly suppressed for the lithiated samples and a bigger concentration of Li in $TiO_2$ reveals a more obvious reduction in the fluctuation range of the cutting force. In addition, we found that the fluctuations are not random and correspond to the changes in the structure of the chips which is not homogeneous with the alternating regions of the crystal and amorphous phase.

To quantify this observation, we counted the number of titanium neighbours for each titanium atom as an order parameter. In the crystal rutile, the number of titanium neighbours for each atom of titanium is 8 inside the sphere with a radius of 4 Å. The density of the amorphous titanium is smaller than that of rutile, therefore the number of neighbours inside the sphere of any determined radius is also smaller on average. Figure 7b shows the snapshots of cutting simulation during cutting of pristine $TiO_2$ and lithiated $TiO_2$. The titanium atoms with 8 titanium neighbours inside the sphere with a radius of 4 Å were marked by blue ("ordered" atoms), and the titanium atoms with less than 8 titanium neighbours were marked by yellow ("disordered" atoms) in Figure 7b. The yellow zone of "disordered" atoms in Figure 7b is believed to be the shear band, which has been observed in the atomistic simulation of shear-band formation during cutting of metallic glasses [52]. Shear band is a narrow region with severe shear strain usually along the primary shear zone. We can find in Figure 7b that four and five approximately periodic intervals of shear bands appear at the cutting chip of $Li_{0.125}TiO_2$ and $Li_{0.25}TiO_2$, respectively. These shear bands around chips of lithiated $TiO_2$ grow from the tool tip until free surface of chips. However, the chip of pristine $TiO_2$ does not show the periodic distribution of shear bands but consists of two big areas of shear bands inside the chip. The irregular formation of shear bands in the chip of pristine $TiO_2$ is believed to arise from the fusion of multiple shear bands [52].

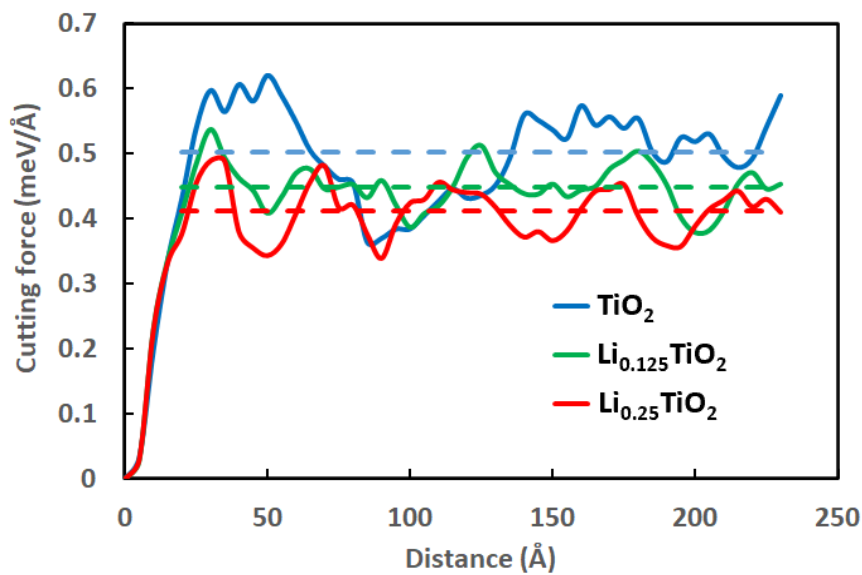

**Figure 8**. The cutting force in $TiO_2$, $Li_{0.125}TiO_2$ and $Li_{0.25}TiO_2$ vs the distance passed by the cutting tool. The dashed line shows an average value the curves in the distance from 20 to 220 Å.

The measured cutting force curves with cutting distance are employed to quantitatively identify the formation of shear bands around chips, as shown in Figure 8. It can be observed that the cutting force increases in the "disordered" yellow region where the cutting is located at the region of shear band and reduces in the "ordered" blue region. The formation of shear band is governed by a critical stress value above which the shear band nucleates and propagates [53]. At the beginning of cutting, the chip deforms by laminar flow with an ordered crystal structure due to a low stress level. The stress gradually increases until the critical stress where the shear band is triggered and propagates towards the chip free surface. In addition, the deformed mode of the chip also alters from laminar to shear localized flow, resulting in the increased cutting force in shear bands. As the stress increases to the fracture stress, the shear band loses efficacy due to stress relaxation with the breakage of the crystal where the chip flow changes into the laminar flow leading to a temporary reduction of cutting load [53]. With the new shear band formation at the tool tip, the cutting force grows and reduces again, which therefore generates the force fluctuation with cutting distance in Figure 8. The shear local flow and laminar flow during chip formation also contribute to the alternate generation of "disordered" and "ordered" crystal structures at the chip in Figure 7b.

Lower fracture stress of lithiated $TiO_2$ also contributes to the periodicity of shear band formation, which supports the stability of the transition from laminar to shear localized flow during chip formation. The distribution of peaks in Figure 8 shows coincidence with the distribution of shear bands around chips in Figure 7b. The cutting curve in $Li_{0.25}TiO_2$ illustrates five distinct peaks with average intervals of approximately 40 Å. For the cutting curve of $Li_{0.125}TiO_2$, four main peaks with average intervals of approximately 50 Å and several secondary peaks with smaller amplitude can be observed. However, blurry periodicity of peaks occurs at the cutting curve of pristine $TiO_2$ where there are two main peaks with bigger peak width compared to lithiated $TiO_2$ and a lot of secondary peaks on the main peaks. These secondary peaks are the result of the formation of secondary shear bands [54], which may be attributed to the merge of main shear bands. These results suggest that greater stability of plastic deformation appears in the cutting process of $Li-TiO_2$, which would favour the improvement of tool wear and surface quality.

## 4 Conclusions

We explored, in a computational study, the possibility of using reversible electrochemical lithiation of titania as a means of non-destructive reversible surface modification to facilitate the cutting of brittle ceramics. We used molecular dynamics simulations with a force field accounting for key mechanistic details of lithiation i.e. the ionization of interstitial Li and formation of $Ti^{3+}$. DFT was performed in the elastic regime on account of a sufficiently small-scale model, and large-scale MD simulations (with up to a million atoms) were performed to compute the effects of lithiation on plastic response and fracture as well as on the cutting force. The conclusions of the study are summarized as follows:

(1) No significant softening in the linear regime was observed at low or room temperature (neither in DFT or MD), but some softening was observed in large-scale MD at elevated temperatures above 1000 K for both pure and lithiated titania.

(2) The elastic modulus and the fracture stress for lithiated rutile are smaller than that for pristine rutile at higher temperatures. The elastic modulus of rutile as well as the fracture stress decrease with increasing temperature.

(3) The fracture stress decreases with Li concentration at all temperatures.

(4) Compared with pristine rutile, the cutting force was found to reduce by about 20% and the oscillations of the cutting force were also found to be smaller during cutting the lithium rutile ($Li_{0.25}TiO_2$).

(5) The lithiated rutile shows better periodicity of shear band formation in the cutting chips, indicating improved deformation stability during cutting process.

The key result of this work is that although at room temperature there was no significant effect on the elastic properties, the effect on the plastic part of the curve and on the fracture, as well as on the cutting force, was noticeable. As similar mechanisms of lithiation (and in general electrochemical doping with alkali or alkali earth atoms) are active in different ceramics, we expect this approach to be useful to facilitate machining with non-destructive reversible modifications of composition of other materials as well, a direction that deserved to be explored in the future.

## Declaration of Competing Interest

We declare that we have no conflict of interest.

## Acknowledgements

This work is supported by the Singapore Ministry of Education, under its Academic Research Funds (Grant No.: MOE-T2EP50120-0010 and MOE-T2EP50220-0010). This work is in part supported by the Ministère des Relations Internationales et la Francophonie du Québec. We also thank Compute Canada on whose servers some of the calculations were performed.